# Emergent lifetime distribution from complex network systems aging

Yimeng Liu[1], Shaobo Sui[1], Dan Lu[1], Rui Peng[2], Mingyang Bai[1*], Daqing Li[1]

[1] Department of Reliability and Systems Engineering, Beihang University, Beijing 100083, China

[2] College of Economics and Management, Beijing University of Technology, Beijing 100124, China

* Corresponding author: baimingyang@buaa.edu.cn

**Abstract**

Most theoretical analysis for lifetime distribution explains origins of specific distribution based on independent failure. We develop one unified framework encompassing different kinds of lifetime distribution for failure coupling system. For typical complex networks, we found that three types of system lifetime distributions are emerged shaped by system size and failure coupling strength. When the failure coupling strength $\phi$ dominates, systems exhibit a cascade failure mode, the system lifetime following an exponential distribution as a series system due to long-range correlation. When the system size $N$ dominates, systems exhibit wear-out failure mode, the system lifetime following the Gompertz model as a parallel system due to short-range correlation. When $N$ and $\phi$ have comparable impact on the system, the system lifetime follows a modified Weibull distribution. We find the critical failure coupling strength and critical system size which are helpful to identify the failure mode switch point of the system. Besides, we provide rigorous theoretical analysis for emerged lifetime distribution. We reveal the microscopic mechanism of system lifetime distribution switch pattern by analyzing the competence between correlation length and network diameter. Finally, we verify our conclusions in real networks. Our study will help understand the lifetime origin of complex systems and design highly reliable systems.

# 1.Introduction

Understanding system lifetime distribution is crucial for designing reliability systems. For a degraded system (the mortality doesn't decrease over time), it can be divided into three situations: the mortality is constant, the mortality increases power law with aging, and the mortality increases exponentially with aging. When the system mortality is constant with time, its lifetime distribution follows the exponential distribution. Exponential distribution is widely used to describe the lifetime distribution of memoryless systems such as electronic components and light bulbs [1]. Drenick's theorem shows that when the series system is sufficiently large, the lifetime distribution of the system tends to follow exponential distribution[2]. When the system mortality increases with time as a power law, its lifetime distribution follows Weibull distribution [3]. Weibull distribution is widely used in lifetime distribution modeling of various types of machinery and other engineering products [4]. When the system mortality increases exponentially with time, its lifetime distribution is described by the Gompertz model [5]. Gompertz's law was discovered in the lifetime distribution of humans and many other animals [6-8], which is widely used in analysis of medical and insurance fields [9-10]. A series of theoretical proofs model system as a large parallel system with amount of redundancy [11-14] and derive Gompertz's law by extreme value theory [15].

But these theories are limited to the assumption that components are uncoupled. Nevertheless, failure correlation between components has been found to exist widely in complex systems [17-20][32-33]. For obtaining lifetime distribution of system with failure coupling, some models have been developed, including universal generating functions [21], reliability block diagrams [22], fault trees [23], decision diagrams [24], Bayesian networks [25] and so on. While the complexity of deriving lifetime distribution by these methods quickly increases when system size grows [39]. To overcome this difficulty, simulation method is widely used for obtaining the system lifetime distribution [16]. However, these methods usually calculate lifetime distribution for different systems in a case-by-case manner, which hinder investigating the roots of different lifetime distributions. There is a lack of one unified framework that reveals the common mechanisms of various lifetime distributions across different systems.

In this study, starting from the Marshall model [26], we construct an Ising-like simplified model. In this model, the failure rate of system components increases with the failure of its coupled components. For four typical network topologies, we analyze how the system lifetime distribution type switches with the system size $N$ and the failure coupling strength $\phi$. We find three types of lifetime distribution emerged from the competence between the system size and failure coupling strength, including Gompertz model, modified Weibull distribution and exponential distribution. The emergence of three lifetime distributions can be fully understood by the relative magnitude between network diameter $D$, which increases with system size $N$, and the correlation length $\xi$, which increases with $\phi$. When the failure coupling strength $\phi$ dominates, correlation length $\xi \gg D$. Then we rigorously analyze the transition between different lifetime distribution. Specially, we observe stable failure propagation and derive a fundamental equation of generalized thermodynamic for large failure-coupled system. Finally, we calculate the lifetime distribution of real networks to verify our conclusions.

## 2. Failure coupling model construction

We derive our model from Marshall model, which gives the lifetime joint distribution of components in system to describe coupled failure. In Marshall model, the $N$ components lifetime $x_1, x_2, \ldots, x_N$ joint distribution of satisfies

$$\bar{F}(x_1, x_2, \ldots, x_N) = P\{X_1 > x_1, X_2 > x_2 \ldots, X_N > x_N\}$$
$$= \exp[-\sum_i r_i x_i - \sum_{i<j} r_{ij} \max(x_i, x_j) - \sum_{i<j<k} r_{ijk} \max(x_i, x_j, x_k) \qquad (1)$$
$$- \cdots - r_{12\ldots N} \max(x_1, x_2, \ldots, x_N)],$$

where failure coupling is represented by the coupled term like $r_{ijk}\max(x_i, x_j, x_k)$. In this study, we consider a simplified model for failure coupling network. Neglecting higher-order terms in Marshall model, we set $r_i = \beta$ for first-order term and $r_{ij} = \beta\phi A_{ij}$ for second-order term. We assume each component with same $r_i = \beta$. Failure coupling relationship is represented by the symmetric unweighted adjacent $A_{ij}$ containing only 1 and 0. Besides we introduce failure coupling strength $\phi$ to quantify the relative magnitude of coupled failure to independent failure. Thus, in our model, the joint probability density function of each component lifetime $p(x_1, x_2, \ldots, x_N)$ satisfy

$$p(x_1, x_2, \ldots, x_N) \propto \exp\{-\beta[\sum_i x_i + \phi \sum_{i<j} A_{ij} \max(x_i, x_j)]\}. \qquad (2)$$

Our model has a similar form to the Ising model [27], which implies our model may also contain rich behavior of failure dependent systems. Different from Ising model, the domain of variable $x_i$ is positive real number in our model instead of discrete set $\{+1, -1\}$.

According to the Eq.(2), the failure rate of healthy component $i$ at time $t$ is

$$\lambda_i(t) = \beta(1 + \sum_j A_{ij} d_j(t) \times \phi), \qquad (3)$$

where $d_j(t)= 0$ represents that component $j$ is still alive at time $t$. $d_j(t)= 1$ represents that component $j$ has failed at time $t$. See SI for details. Here, $\beta$ is the failure rate of each component when it has no dead neighbor. In our model, we consider both primary fault and depend fault (as shown in Fig.1). When there is no failure coupling between the components, the failure of the network components is independent of each other. When there is a failure coupling relationship between components, the failure of one component will lead to an increase in the failure rate of the components depending on it. The failure of one component will lead an increase in the failure rate of its neighbor. The increased failure rate is $\phi$ times the initial failure rate. When $\phi = 0$, each component fails independently. When $\phi \to \infty$, the failure mode of system is dominated by failure propagation from dead nodes dominate.

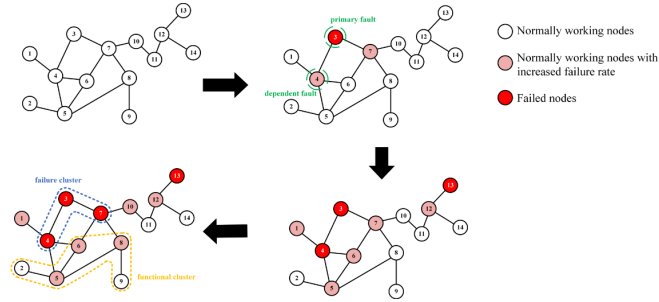

**FIG. 1 Failure propagation diagram of the network failure coupling model**

**3. System lifetime distribution analysis with different failure coupling strength or system size**

In the following, we focus on the two parameters of system size and coupling strength to analyze the system lifetime distribution properties in typical networks based on Eq.(3) and Gillespie algorithm [28] (see SI for the process). We use 2D square lattice network, ER network, WS network (rewiring probability is 0.01) and BA network these four typical complex networks as examples to analyze [43], where the average degree is all set to 4. We define that system is dead when the number of failed nodes in the system exceeds a certain proportion, $p_c$. The value of $p_c$ depends on the situation and could be set as the percolation threshold [29] ($p_c = 0.1$ in our study). The KL divergence between three models (exponential model, Gompertz model and modified Weibull model) and the empirical data of system lifetime was calculated to determine the fitting effect of the model. The failure rate and reliability function of models are shown in Appendix A. The KL divergence calculation method is shown in SI. According to Occam's razor principle, we will select the model with the fewest parameters among three models (model parameters are obtained by maximum likelihood estimation) within an acceptable range (KL divergence <0.2) as the system lifetime distribution model.

**3.1 Lifetime distribution analysis with different failure coupling strength**

First, we study the system lifetime distribution properties under different coupling strengths $\emptyset$. Taking the 2D square lattice network as an example, we find that as the coupling strength increases, the model of system lifetime distribution switches from the Gompertz model to the Modified Weibull distribution, and finally to the exponential distribution (as shown in Figure 2). Here, the lifetime samples are normalized to ensure their average value is 1, which does not affect the type of lifetime distribution. When there is no coupled failure ($\emptyset=0$), the fitting deviations of the modified Weibull distribution and the Gompertz model are both within the acceptable range, which can describe the lifetime distribution well (as shown in Fig 2(a) and (b)). Considering that the Gompertz Model has fewer parameters, we believe that in this case the Gompertz Model is more suitable for describing the system lifetime distribution. This can be understood as when the coupling strength is weak, the system aging behaves as a wear-out failure mode and dies only when the damage gradually reaches the threshold. The speed of damage remains stable during damage accumulation. Therefore, the lifetime distribution is relatively concentrated which is consistent with the Gompertz model. When $\emptyset=10^4$, only the modified Weibull distribution can better describe the lifetime distribution (as shown in Fig 2(c) and (d)). When $\emptyset=10^6$, although all three models can describe the lifetime distribution well, the exponential distribution with the fewest parameters is more suitable to model the lifetime distribution (as shown in Fig 2(e) and (f)). At this time, it can be understood that when the coupling strength is strong enough, the system aging behaves as a cascading failure. Once a system component fails, the failure will propagate quickly and cause the system to collapse rapidly. Therefore, the probability of the system dying in unit time can be approximated by the probability $N\beta$ that the system going from being completely healthy to having components failed in the system within the unit time, so the system lifetime follows an exponential distribution. This result was also found in ER network, small-world network, and BA network (see SI).

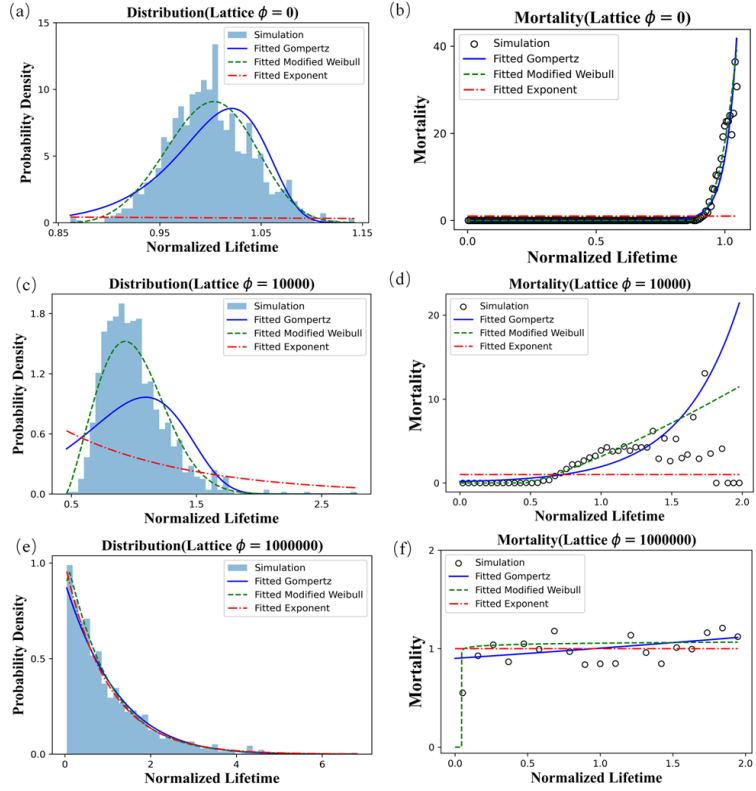

**Fig. 2 The relationship between the lifetime distribution and mortality of the system over time with different failure coupling strength in 2D square lattice network.** (a), (c) and (e) are the lifetime distribution of the system when the coupling strength is 0, $10^4$, and $10^6$ respectively. (b), (d) and (f) are the system mortality over time when the coupling strength is 0, $10^4$, and $10^6$ respectively (see SI for the mortality calculation method). System size is 6400. The blue solid, green dashed and red solid dotted fitting curves in the figure are the Gompertz model, modified Weibull distribution and exponential distribution respectively. The fitting effect is measured by calculating the KL divergence between the fitting function and the simulation data. KL divergence values in each condition shown in Table B1 in Appendix B.

### 3.2 Lifetime distribution analysis with different system size

When we study the system lifetime distribution properties under different system sizes, we find that as the system size increases, the lifetime distribution of the system switches from exponential distribution (as shown in Figure 3(a)) to modified Weibull distribution (as shown in Figure 3(b)), and finally to Gompertz model (as shown in Figure 3(c)). The mortality rate switches from exponential increasing (as shown in Figure 3(b)) to power law increasing (as shown in Figure 3(d)), and finally to being constant (as shown in Figure 3(f)). This switch can be understood that when $N$ is sufficiently small ($N$=1 in the limit), the system lifetime tends to be similar as one components lifetime. In this case, the lifetime distribution of the system tends to follow the exponential distribution. The increase in system size $N$ will increase system redundancy, which is widely used for enhancing reliability in engineering, thereby reducing the fluctuation of system lifetime. When $N$ is sufficiently large, system lifetime distribution may be highly concentrated so that the mortality rapidly increases at an exponential rate near the expected value of the life distribution, showing Gompertz's law. This result is also found in ER network, small world network and BA network (see SI). Above all, it is clear that both system size and coupling strength can affect the type of system life distribution with the opposite trend.

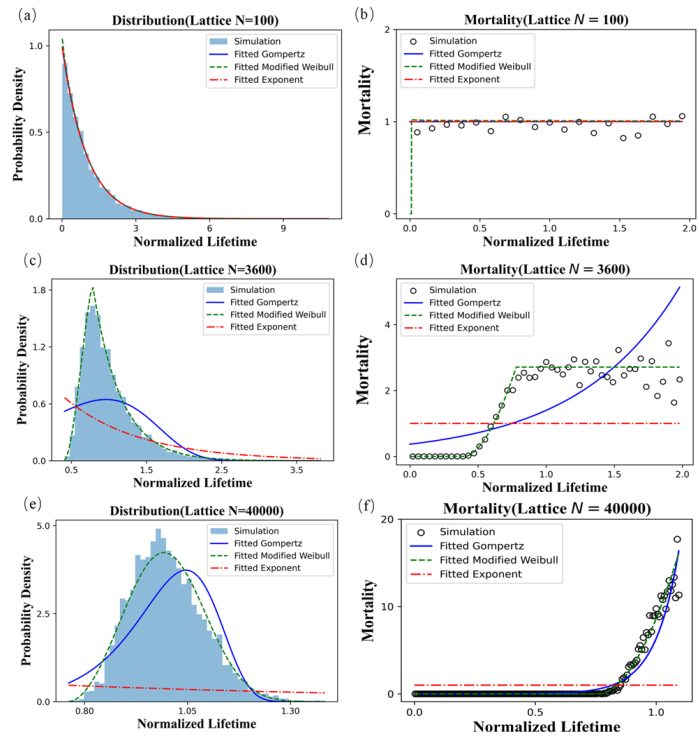

**Fig. 3 The relationship between the lifetime distribution and mortality of the system over time with different system sizes in 2D square lattice network**. (a), (c) and (e) are the lifetime distribution of the system when the system size is 100, 3600 and 40000 respectively. (b), (d) and (f) are system mortality over time when the system size is 100, 3600, and 40000, respectively. Coupling strength is $10^4$. The blue solid, green dashed and red solid dotted fitting curves in the figure are the

Gompertz model, modified Weibull distribution and exponential distribution respectively. The fitting effect is measured by calculating the KL divergence between the fitting function and the simulation data. KL divergence values in each condition shown in Table B2 in Appendix B.

**4. Lifetime distribution emergent from competence between coupling and redundancy**

To understand how failure coupling strength and system size collectively affect system lifetime distribution, we study the suitable lifetime distribution types under different system sizes $N$ and coupling strengths $\phi$. The competence between failure coupling, which induce system vulnerability, and system size, which increases system redundancy, is revealed. Based on the KL divergence, we find that the four-parameter modified Weibull model can fit the system lifetime distribution well in each parameter area in 2D square lattice network since the KL divergence is always small (Fig. 4(b)). The 2-parameter Gompertz model and exponential distribution are special cases of modified Weibull model after degeneration when the system size $N$ or the coupling strength $\phi$ is sufficiently large (Fig. 4(a) and 4(c)). Thus, the system lifetime distribution type heatmap is divided into three areas (as shown in Figure 4(d)). $N$ and $\phi$ show the opposite tendency. The impact of increasing coupling strength $\phi$ is similar to the impact of decreasing system size $N$ on the system lifetime distribution. This is because when the failure coupling strength increases, multiple strongly coupled components can be viewed as one component effectively, namely the system size is decreasing. When the failure coupling strength is sufficiently strong, the entire system can be regarded as one component.

Therefore, three types of lifetime distribution emerged from the competence between $N$ and $\phi$: (1) When the system size $N$ dominates, the system failure process appears as wear-out mode and the lifetime distribution follows the Gompertz Model as a parallel system. (2) When the failure coupling strength $\phi$ dominates $N$, the system failure process appears as cascading failure and the lifetime distribution follows exponential distribution as a series system. (1) When $N$ and $\phi$ have comparable impact on the system, system failure mode is a combination of failure propagation and wear-out, which must be described by Modified Weibull model with more parameters. This result was also found in ER network, small-world network, and BA network (see SI).

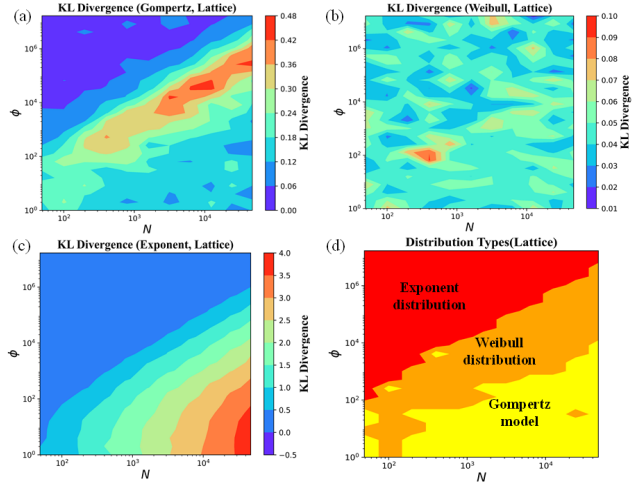

**Fig. 4 Phase diagram of network lifetime distribution in 2D square lattice network.** (a)-(c) KL Divergence between empirical data and Gompertz model, modified Weibull distribution model and Exponent distribution model under different coupling strengths and different system sizes, respectively. (d) 2D square lattice network lifetime distribution types under different coupling strengths and different system sizes.

In particular, we find that KL Divergence between empirical data and Gompertz model is small when system size $N$ dominates or failure coupling strength $\phi$ dominates. When $N$ and $\phi$ have comparable impact, KL Divergence between empirical data and Gompertz model is large. This comes from the fact that when the influence of $\phi$ dominates, as shown in Figure 4(d), the system lifetime distribution follows exponential distribution(see Fig. 2(e)-2(f), Fig. 3(a) and 3(b)). At this time, the Gompertz model can also describe system lifetime distribution well by setting the model parameter $B$ (see Appendix A) to 0. Therefore, the KL divergence between the Gompertz model and the empirical data is small at this time and the system exhibits a cascading failure mode. When the influence of $N$ dominates, the system mortality rate increases rapidly and the system lifetime distribution is best described by the Gompertz model. Therefore, the KL divergence between the Gompertz model and the empirical data is small at this time and the system exhibits a wear-out failure mode. When $N$ and $\phi$ have comparable impact, the Gompertz model can't describe the system lifetime distribution so KL divergence is large and the system exhibits a mixed failure mode. Therefore, we believe the KL divergence of Gompertz and empirical distribution will help us identify the critical points at which the system switches to different failure modes.

Here we conducted a detailed analysis of the variation in KL divergence under fixed system size or fix failure coupling strength. As an example in 2D square lattice network, for a given system network structure and system size $N$, as the coupling strength $\phi$ increases, the KL divergence first increases and then decreases. The maximum point of the KL divergence, denoted as $\phi_c$, is the critical point of the system failure coupling strength (as shown in Fig. 5(a)). When the fault coupling strength $\phi \ll \phi_c$, the system is weakly coupled and exhibits the wear-out failure mode. The lifetime distribution follows Gompetz model. When the fault coupling $\phi \gg \phi_c$, the system is strongly coupled. The failure of a component will occur rapidly, leading to the failure of the entire network. The lifetime distribution follows the exponential distribution. When $\phi \sim \phi_c$, the system failure mode is a mixture of two failure modes. Similarly, for a given system network structure and failure coupling strength, as the system size $N$ increases, the KL divergence also first increases and then decreases. The maximum point of the KL divergence, denoted as t $N_c$, is the critical point of the system size (as shown in Fig. 5(b)). When fault coupling $N \ll N_c$, the system has little redundancy and the lifetime distribution follows exponential distribution. As the number of system components increase, when the system scale is $N \gg N_c$, the system has enough redundancy and the failure of a few components will not immediately cause system to collapse. At this time, the system exhibits a wear-out failure mode and the system lifetime distribution exhibits the wear-out failure mode. When $N \sim N_c$, the system failure mode is a mixture of two failure modes. This result was also found in ER network, small-world network, and BA network (see SI).

This pattern can help the reliability design of large systems. When the structure and failure coupling strength of the interested system are known, we can design the system size $N$ to be much larger than the critical point $N_c$ to avoid the system rapidly collapsing due to failure of a few components by obtaining the system size critical point $N_c$. Besides, for a network system with a given structure and system size, we can take measures to decrease the failure coupling strength below $\phi_c$ to improving the reliability of the system by obtaining the failure coupling strength critical point $\phi_c$.

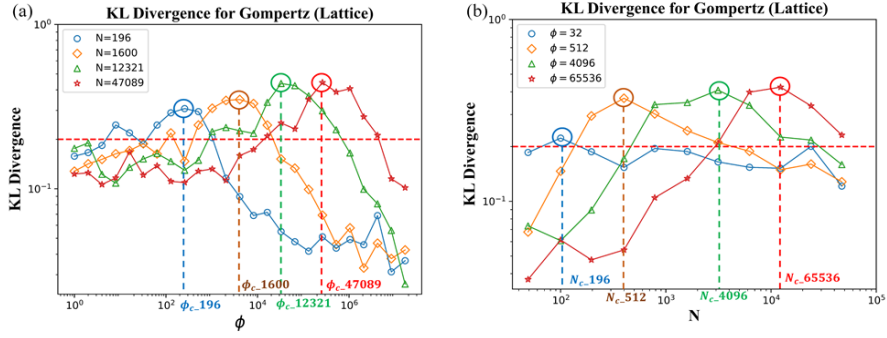

**Fig. 5 The variation of KL divergence values between the Gompertz model and empirical distribution under fixed system size or fix failure coupling strength as an example in 2D square lattice network.** (a) The variation of KL divergence values between the Gompertz model and empirical distribution with failure coupling strength when the system size is 196, 1600, 12321 and 47089. (b) The variation of KL divergence values between the Gompertz model and empirical distribution with system size when the failure coupling strength is 32, 512, 4096 and 65536.

## 5. Theoretical analysis for lifetime distribution switch pattern

Switch of three system lifetime distribution types can be well explained by our theory. Below, we derive the lifetime distribution of the system theoretically under two extreme cases: failure coupling strength $\phi$ dominates (N=const, $\phi \to \infty$) or system size $N$ dominates (N→∞, $\phi$=const). First, we derive the equation for the system lifetime. The time interval from the moment when $n^{th}$ node fails to the moment when $(n+1)^{th}$ node fails in the system is denoted as $T_n$. If the system dies when $p_c$ proportion of nodes failed, the total system lifetime is

$$T_{sys} = \sum_{n=0}^{[Np_c]-1} T_n, \tag{4}$$

where $[Np_c]$ represents the largest integer that is not greater than $Np_c$. To get the probability distribution for $T_n$, we calculate the transition probability from state of $n$ failed nodes to state of $n + 1$ failed nodes, which is the sum of failure rate for $N$-$n$ healthy nodes

$$\lambda_n = \beta(N - n) + \beta\phi\omega(n, N, \phi), \tag{5}$$

where the first term is the initial failure rate of $N$-$n$ healthy nodes due to their own failure represented by $r_i = \beta$. The second term is the increased failure rate of $N$-$n$ healthy nodes caused by $n$ failed nodes represented by $r_{ij} = \phi A_{ij}$. $\omega(n, N, \phi)$ is the number of edges between the failed nodes and the healthy nodes. Thus $T_n$ follows the exponential distribution with parameter $\lambda_n$. Due to the Markov properties of the failure process, $T_n$(n=0,1,2,…,$N-1$) are independent from each other. The moment generating function of the system lifetime $T_{sys}$ satisfies

$$E[e^{tT_{sys}}] = \prod_{n=0}^{[Np_c]-1} E[e^{tT_n}] = \prod_{n=0}^{[Np_c]-1} \frac{1}{1-\frac{t}{\lambda_n}}. \tag{6}$$

This generating function describes all the properties of the system life distribution.

When the impact of coupling strength $\phi$ far exceeds the impact of system size $N$, it can be expressed as N=const, $\phi \to \infty$. Currently, $\lambda_0 = \beta N$ is a constant, and $\lambda_n \to \infty$ for $n = 1,2,\ldots[Np_c]-1$. The moment generating function of the system lifetime $T_{sys}$ is

$$\lim_{\phi \to \infty} E[e^{tT_{sys}}] = \lim_{\phi \to \infty} \prod_{n=0}^{[Np_c]-1} E[e^{tT_n}] = \frac{1}{1-\frac{t}{\lambda_0}} \prod_{n=1}^{[Np_c]-1} \lim_{\phi \to \infty} \frac{1}{1-\frac{t}{\lambda_n}} = \frac{1}{1-\frac{t}{\lambda_0}}, \tag{7}$$

by using $\lim_{\phi \to \infty} \frac{1}{1-\frac{t}{\lambda_n}} = \lim_{\lambda_n \to \infty} \frac{1}{1-\frac{t}{\lambda_n}} = 1$ for $n = 1,2,\ldots[Np_c]-1$. At this situation, the system lifetime converges to an exponential distribution with parameters $\lambda_0 = N\beta$ under $\phi \to \infty$.

When the impact of system size $N$ far exceeds the impact of coupling strength $\phi$, it can be expressed as N$\to\infty$, $\phi$ =const. Unlike the situation of $\phi\to\infty$ where we only need to pay attention to the initial failure, the failure propagation process of the system must be clearly analyzed for the life distribution of the system (represented as $\omega(n,N,\phi)$), which is hindered by the complex coupling network structure. Here we applied thermodynamics to overcome this difficulty. For a large system consist of many independent small parts, thermodynamics[30] of extensive variables can be constructed [36]. According to central limit theorem, the average of additive variables for different small parts will become a well-defined macroscopic variable. For a large network, different subsystem can be seem almost independent. The number of edges between failed components and healthy components is selected as the interested extensive variable. For a fixed network structure and coupling strength $\phi$, we can define a intensive variable for the system as

$$\bar{\omega}(x,\phi) = \lim_{N \to \infty} \frac{\omega(xN,N,\phi)}{N}, \tag{8}$$

where $x$ is the proportion of failed nodes. We verified that this limit exists for different network structures and coupling strengths (see appendix C).

After the failure evolution process of the system under the thermodynamic limit is obtained, we prove that the probability density function $P(t_{sys})$ of the system lifetime $T_{sys}$ under the thermodynamic limit satisfies

$$P(t_{sys}) \asymp e^{-NI(t_{sys})}, \tag{9}$$

where $T_{sys}$ follows large deviation principle. $I(t_{sys})$ is rate function, corresponding to the entropy function in statistical physics [31]. We will prove the above relationship below and give an equation for $I(t_{sys})$. First, substituting Eq.(4-6), the expression of scaled generating function of $T_{sys}$ under the thermodynamic limit is

$$H(k) = \lim_{N \to \infty} \frac{\ln(E[e^{NkT_{sys}}])}{N}$$
$$= \lim_{N \to \infty} \sum_{n=0}^{[Np_c]-1} -\ln\left(1 - \frac{k}{\beta(1-\frac{n}{N}+\phi\bar{\omega}(\frac{n}{N},\phi)+o(1))}\right)\frac{1}{N} \tag{10}$$
$$= \int_0^{p_c} -\ln\left(1 - \frac{k}{\beta(1-x+\phi\bar{\omega}(x,\phi))}\right)dx,$$

where the domain of $k$ is $(-\infty, \min_{x \in [0,p_c]} \beta(1-x+\phi\bar{\omega}(x,\phi)))$ and the range is $(-\infty, +\infty)$. Since $H(t)$ is a convex function about $t$, $I(a)$ is the Legendre-Fenchel transform of $H(t)$ using the Gartner-Ellis Theorem [31] as

$$I(t_{sys}) = \sup_{k \in \mathbb{R}} \{kt_{sys} + \int_0^{p_c} \ln\left(1 - \frac{k}{\beta(1-x+\phi\bar{\omega}(x,\phi))}\right)dx\}. \tag{11}$$

Thus, the asymptotic lifetime distribution Eq.(9) in thermodynamical limit ($N \to \infty$) is derived. Notably, rate function $I$ corresponds to entropy function. Expressing $I = I(t_{sys}, \phi, \bar{\omega})$ as the function of macroscopic variables, Eq.(11) is essentially a fundamental equation of generalized thermodynamics for large failure-coupled system, which will induce generalized thermodynamic relationships [36]. Based on scaled generating function, it can be seen that the coefficient of variation of $T_{sys}$ is $CV[T_{sys}] = \frac{\sqrt{Var[T_{sys}]}}{E[T_{sys}]}$ which satisfies that

$$\lim_{N \to \infty} \frac{CV[T_{sys}]}{\frac{1}{\sqrt{N}}} = \frac{\sqrt{H''(0)}}{H'(0)}. \tag{12}$$

Therefore, CV[$T_{sys}$] decays at the rat of $\frac{1}{\sqrt{N}}$ with increasing system size $N$, which is consistent with the simulation results (as shown in Fig. 6). Given assessing the premature failure risk is needed in reliability engineering, while getting enough lifetime data is expensive for large system. With this scaling law, we can extrapolate the coefficient of variation of large system lifetime from easily accessible lifetime data of small system.

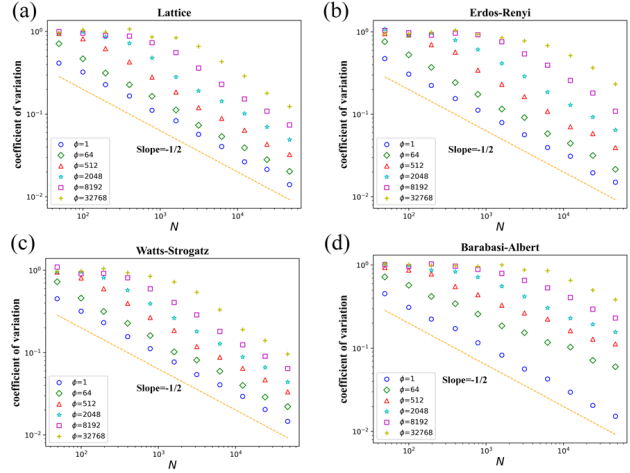

**Fig. 6 Relationship between system size $N$ and coefficient of variation.** (a) 2D square lattice network structure. (b) ER network structure. (c) Small world network structure. (d) BA network structure.

The scaling decaying suggests failure coupling systems with finite $\phi$ are self-averaging, which comes from the correlation length $\xi$ is significantly smaller than the network diameter $D$ in disorder systems [34-35]. The concept of self-averaging posits that the fluctuations within a system vanish in strength as the size of the system increases. Due to self-averaging property, system lifetime follows peaked distribution in thermodynamical limit ($N \to \infty$), which induces mortality increases exponentially following Gompertz model. As $\phi$ increases, the increasing correlation length $\xi$ will break the self-averaging [37-38], which can be compensated by increasing $N$. Thus, the opposite effect of $N$ and $\phi$ could be explained from perspective of self-averaging. The correlation length is characterized by the correlation function. In systems with a short correlation length, the correlation between nodes quickly diminishes as the distance between nodes increases. Conversely, in systems with a long correlation length, the correlation remains substantial even for pairs of remote nodes.

To understand the microscopic mechanism behind lifetime distribution switch pattern, we further analyzing the correlation function in four typical network structure. We calculate the Pearson correlation coefficient of lifetime for all pairs of nodes. The correlation function is calculated as the average correlation among pairs of nodes share same distance (the shortest path length between this pair of nodes). Thus we analyze how the correlation function decreases with normalized distance (the ratio of the shortest path length to the network diameter $D$). We find that when the network system structure and system size $N$ are fixed, the increase in failure coupling strength $\phi$ leads to slower decline of the correlation (as shown in Figure 7(a)), which corresponds to long correlation length $\xi$. The self-averaging property breaks. For system with long correlation length $\xi$, a failure of one component will quickly propagate to the entire network system. When the network system structure and failure coupling strength $\phi$ are fixed, the increase in system size $N$ leads an increase in the network diameter, causing the decline of the ratio of correlation length $\xi$ to system size (as shown in Figure 7(b)) and emergence of self-averaging property. The increasing redundancy will prevent system from immediately collapse due to the failure of a few components. Thus the different result between networks can be understood by their structure. With same failure coupling strength $\phi$, the correlation decreases more smaller in BA networks and ER networks (as shown in Figure 7(c-d)) due to their small network diameter [43] .This explains why the coefficient of variation decreases more slower (as shown in Figure 6(c-d))..

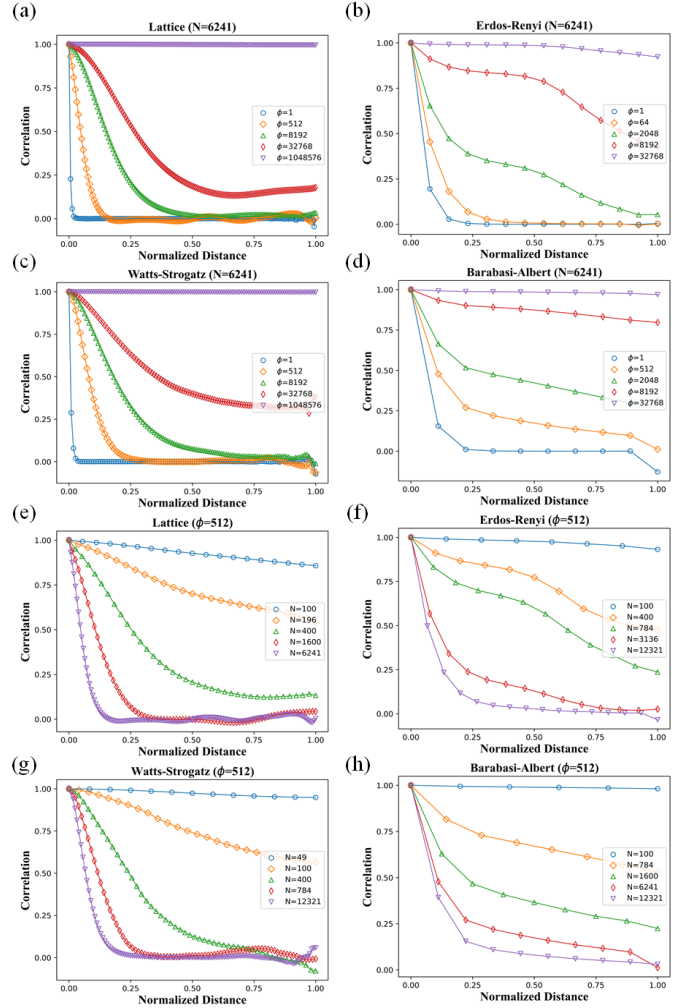

**Fig. 7 The variation of correlation function with normalized distance between pairs of components.** (a)-(d) The variation of correlation function with normalized distance between pairs of components with fix system size (*N*=6241) in 2D square lattice network, ER network, small world network and BA network, respectively. (f)-(h) The variation of correlation function with normalized distance between pairs of components with fix system failure coupling strength (*ϕ*=512) in 2D square lattice network, ER network, small world network and BA network, respectively.

## 6. Case study

Finally, we further verified the above results in real network. A concrete example network we analyze is the western states power grid of the United States showing below [40]. We find that as the coupling strength increases, the switch process of system lifetime distribution types is consistent with the previous results. For a real power network with fixed structure and network size, as the coupling strength increases, system lifetime distribution switches from the Gompertz model to the Modified Weibull distribution, and finally to the exponential distribution (shown in Fig. 8(a)). The mortality rate of system witches from exponential increasing to power law increasing, and finally to being constant (shown in Fig. 8(b)). The coupling strength $\phi$ is divided into three regions (shown in Fig. 8(c)). In this case, we can find the critical point of failure coupling strength $\phi_c \approx 10^4$. To understand the microscopic mechanism behind switch between different lifetime distribution, we also analyze how the Pearson correlation of pairs of nodes lifetime vary with normalized distance. As shown in Fig. 8(d), with increasing $\phi$, correlation decreases slower (Fig. 8(d)), which corresponds to a larger correlation length $\xi$. When correlation length $\xi \ll D(\phi = 1$ in Fig. 8(d)), lifetime distribution can be better characterized by the 2-parameter Gompertz model. When correlation length $\xi \gg D(\phi = 10^6$ in Fig. 8(d)), highly-vulnerable system lifetime can be better characterized by the 1-parameter exponential distribution. With correlation length $\xi \sim D(\phi = 10^4$ in Fig. 8(d)), system lifetime must be described by the modified Weibull distribution with more parameters. In addition, we also obtained similar results in Gnutella P2P network [41] and biological gene network [42] (see Appendix D).

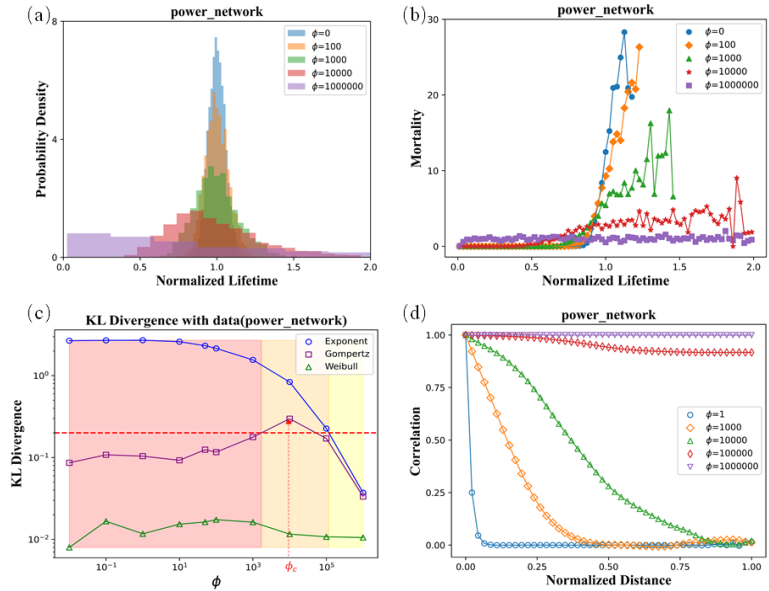

**FIG. 8 The switch process analysis of system lifetime distribution types in real power network.** (a) The lifetime distribution of the system when the coupling strength is 0, $10^2$, $10^3$, $10^4$, and $10^6$. (b)The system mortality over time when the coupling strength is 0, $10^2$, $10^3$, $10^4$, and $10^6$. (c) KL

divergence values between the three fitting distributions and the empirical distribution. System in light red (left) zone follows Gompertz model, in light orange (middle) zone follows Weibull model and in light yellow (right) zone follows exponent model. (d) The correlation between nodes decreases with normalized distance when the coupling strength is $0, 10^3, 10^4, 10^5$ and $10^6$. The pair correlation is the Pearson of lifetime of node in 100 simulations. The power gird network has 4,942 nodes with 6,594 edges.

## 7. Conclusions

In conclusion, we construct a simplified model for failure coupling systems to study the emergence of different lifetime distribution. We find that three distribution emerged under the influence of system size $N$ and failure coupled strength $\phi$. The switch between different lifetime distributions can be well understood by the competence between network diameter $D$, which increases with $N$, and correlation length $\xi$, which increases with coupling strength $\phi$. The relationship between correlation length and failure mode is also verified by simulation. (1)When the failure coupling strength $\phi$ dominates($\xi \gg D$), system exhibits a cascading failure mode. Due to constant initial failure rate, lifetime follows exponential distribution. (2)When the system size $N$ dominates($\xi \ll D$), system demonstrate wear-out failure mode. Due to self-averaging property, lifetime distribution is peaked and follow Gompertz model. (3) When N and $\emptyset$ have comparable impact on the system($\xi \sim D$), without separation of scales to simplify analysis, lifetime distribution must be described by modified Weibull model with more parameters. Based on the KL Divergence between empirical data and Gompertz model, we identify the critical failure coupling strength and critical system size, which are helpful to determine when system failure mode switches. Besides, we rigorously derive the lifetime distribution in the strong coupling limit and thermodynamical limit. Specially, we find the failure propagation pattern can be accurately described in the thermodynamical limit. The fundamental equation of generalized thermodynamics we derived extends thermodynamics to large failure-coupled system. Finally, we verify our conclusions in a real power network, Gnutella P2P network and biological gene network. Our study will help understand the lifetime origin of complex systems and design highly reliable systems.

**Declaration of competing interest**

**Data availability**

The data of the real networks are public data. They can be found in the links in the corresponding references

**Acknowledgments**

This work is supported by the National Natural Science Foundation of China (Grants 72225012, 71822101), the Fundamental Research Funds for the Central Universities.

**Appendix A: Calculation method for failure rate, reliability function and system mortality**

Note probability density function of system lifetime T is $p(t)$. The failure rate and reliability functions of the three models are shown in Table 1. The Reliability function ($R(t) = P(T > t)$) represents the probability that the system is alive at time $x$. The Failure rate ($\lambda(t) = \frac{P(t+dt>T>t)}{P(T>t)dt} = \frac{p(t)}{R(t)}$) is the probability of system dies in per time. The probability density function can be obtained by multiplying the two, namely $p(t) = R(t)\lambda(t)$.

Table 1 Failure rate and reliability functions of fitting models

| Distribution | Failure rate $\lambda(t)$ | Reliability function $R(t)$ | $p(t)$ |
|---|---|---|---|
| Modified Weibull | $0, 0 < t \leq a$ <br> $\frac{c}{b}(\frac{t-a}{b})^{c-1}, a < t < d$ <br> $\frac{c}{b}(\frac{d-a}{b})^{c-1}, t \geq d$ | $1, 0 < t \leq a$ <br> $e^{-(\frac{t-a}{b})^c}, a < t < d$ <br> $e^{-(\frac{d-a}{b})^c - \frac{c}{b}(\frac{d-a}{b})^{c-1}(t-d)}, x \geq d$ | $R(t)\lambda(t)$ |
| Exponential | $\frac{1}{\theta}, t > 0$ | $e^{-\frac{t}{\theta}}, t > 0$ | |
| Gompertz | $Be^{At}, t > 0$ | $e^{\frac{B}{A}[1-\exp(At)]}, t > 0$ | |

Here, the modified Weibull distribution is an extension of the commonly used Weibull model. It degenerates to the normal three-parameter Weibull distribution when $d \to \infty$. The parameter $d$ is introduced to describe the phenomenon that system failure rate first increase and then stay the same, which we observe in the model. After the introduction of parameter *d*, the modified Weibull model can fit the empirical distribution well over a wide range of parameter ranges (see SI).

**Appendix B: KL divergences of three fitting model and empirical data under different coupling strength and system size**

Table B1 KL divergences of three fitting model and empirical data under different coupling strength

|  | Gompertz model | modified Weibull distribution | exponential distribution |
|---|---|---|---|
| $\phi=0$ | 0.1351 | 0.0387 | 2.8028 |
| $\phi=10^4$ | 0.3785 | 0.0671 | 1.1009 |
| $\phi=10^6$ | 0.0601 | 0.0195 | 0.0664 |

Table B2 KL divergences of three fitting model and empirical data in different system size

|  | Gompertz model | modified Weibull distribution | exponential distribution |
|---|---|---|---|
| N=100 | 0.0219, | 0.0116 | 0.0218 |
| N=3600 | 0.4322 | 0.0107 | 0.8032 |
| N=40000 | 0.1635 | 0.0122 | 2.0286 |

**Appendix C: The relationship between the proportion of failure nodes and the number of edges between failure nodes and healthy node**

To prove that $\omega(n, N, \phi)$ is a linear homogeneous function about $n, N$ (widely used to describe thermodynamic systems) and there exists a limit: $\bar{\omega}(x, \phi) = \lim_{N \to \infty} \frac{\omega(xN, N, \phi)}{N}$ when N is sufficiently large, we calculated the relationship between the proportion of failure nodes and the number of edges between failure nodes and healthy nodes under different coupling strength and network structure. The results are shown in Figure B1. The horizontal coordinate is the proportion of failure nodes, and the vertical coordinate is the ratio of the number of connected edges between failure nodes and healthy nodes to the size of the network. For given size of network, we counted the number of edges connecting failure nodes to healthy nodes when the proportion of failure nodes is $x$ ($x$=0.01,0.02,…,0.99). Ordinate value is obtained by dividing the results by the network size. For different coupling strengths $\phi$ and various network structures, we all found that as the system size $N$ increases, the ratio of the number of connections between failure nodes and healthy nodes to the network size gradually converges to a curve. The limits of theoretical analysis are verified to exist from the perspective of simulation.

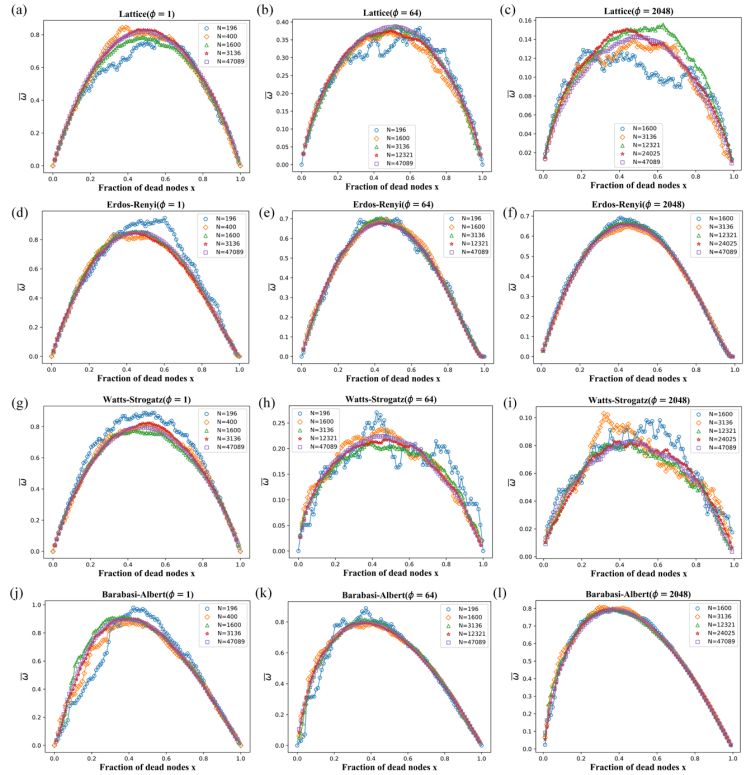

**Fig.C1 Relationship between the proportion of failure nodes and the number of edges between failure nodes and healthy nodes under different network structures and network sizes.** When the couped strength and network structure are fixed, the relationship between the proportion of failure nodes and the number of edges between failure nodes and healthy nodes converges to a curve.

**Appendix D: The results of switch process analysis of system lifetime distribution types in real Gnutella P2P network and biological gene network.**

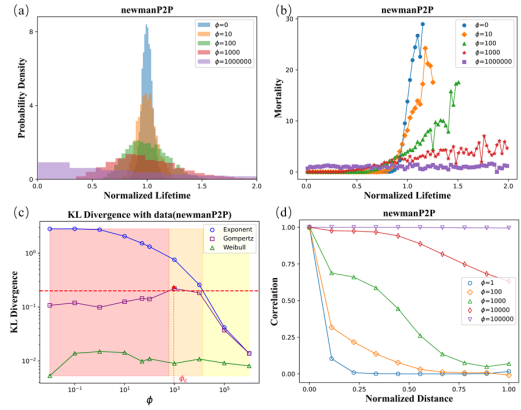

**Fig.D1 The switch process analysis of system lifetime distribution types in Gnutella P2P network.** (a) The lifetime distribution of the system when the coupling strength is 0, 10, $10^2$, $10^3$, and $10^6$. (b)The system mortality over time when the coupling strength is 0, 10, $10^2$, $10^3$, and $10^6$. (c) KL divergence values between the three fitting distributions and the empirical distribution. System in light red (left) zone follows Gompertz model, in light orange (middle) zone follows Weibull model and in light yellow (right) zone follows exponent model. (d) The correlation between nodes decreases with distance when the coupling strength is $0, 10^3, 10^3, 10^4$ and $10^5$. Distance of pair of nodes is defined as their shortest path length. The pair correlation is the Pearson of lifetime of node in 100 simulations. The Gnutella P2P network has 6301 nodes with 20777 edges.

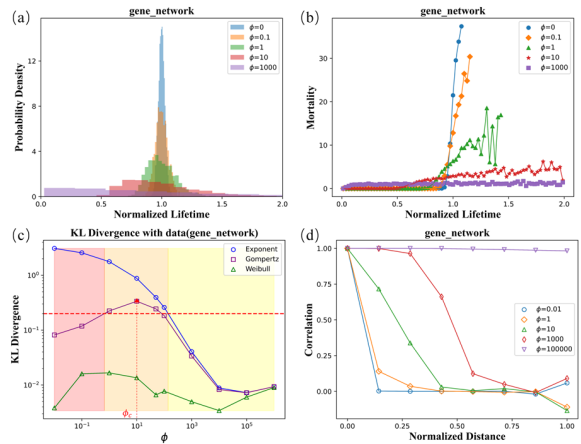

**Fig.D2 The switch process analysis of system lifetime distribution types in real biological gene network.** (a) The lifetime distribution of the system when the coupling strength is 0, 0.1, 1, 10 and $10^3$. (b)The system mortality over time when the coupling strength is 0, 0.1, 1, 10 and $10^3$. (c) KL divergence values between the three fitting distributions and the empirical distribution. System in light red (left) zone follows Gompertz model, in light orange (middle) zone follows Weibull model

and in light yellow (right) zone follows exponent model. (d) The correlation between nodes decreases with distance when the coupling strength is 0.01, 1, 10, $10^3$ and $10^5$. Distance of pair of nodes is defined as their shortest path length. The pair correlation is the Pearson of lifetime of node in 100 simulations. The biological gene network has 14340 nodes with 9041364 edges.

# Supplementary Information for:

# Emergent lifetime distribution from complex network systems aging

## 1. Failure-coupled network model based on Marshall-Orkin Model

The Marshall-Orkin model is a representative model that describes the failure coupled phenomenon through joint lifetime distribution and is suitable for theoretical analysis [39]. In the Marshall-Orkin model, for a system composed of $N$ components, the lifetime of each component $X_i (i = 1,2,\ldots,N)$ is a random variable and their joint distribution satisfies [26]

$$\bar{F}(x_1, x_2, \ldots, x_N) = P\{X_1 > x_1, X_2 > x_2 \ldots, X_N > x_N\}$$
$$= \exp[-\sum_i r_i x_i - \sum_{i<j} r_{ij} \max(x_i, x_j) - \sum_{i<j<k} r_{ijk} \max(x_i, x_j, x_k)$$
$$- \cdots - r_{12\ldots N} \max(x_1, x_2, \ldots, x_N)], \qquad (S1)$$

where the domain is $(0, +\infty)^N$. When all coupling term are set to zero, their joint distribution satisfies

$$\bar{F}(x_1, x_2, \ldots, x_N) = P\{X_1 > x_1, X_2 > x_2 \ldots, X_N > x_N\}$$
$$= \exp(-\sum_i r_i x_i) = \prod_{i=1}^N \exp(-r_i x_i) \qquad (S2)$$

Then

$$P\{X_i > x_i\} = \exp(-r_i x_i) \qquad (S3)$$

Lifetime of each component $i$ satisfy exponential distribution with const failure rate $r_i$. To derive a simplified model of failure-coupled network, we only consider two-component coupling and neglecting higher-order coupling. Thus the joint distribution for component lifetime satisfies

$$\bar{F}(x_1, x_2, \ldots, x_N) = P\{X_1 > x_1, X_2 > x_2 \ldots, X_N > x_N\}$$
$$= \exp[-\sum_i r_i x_i - \sum_{i<j} r_{ij} \max(x_i, x_j)]. \qquad (S4)$$

Below, we derive the joint probability density function $p(x_1, x_2, \ldots, x_N)$ for joint distribution defined by Eq.(S4). According to the definition of joint probability density function, we have

$$\bar{F}(x_1, x_2, \ldots, x_N) = \int_{x_1}^{+\infty} dy_1 \int_{x_2}^{+\infty} dy_2 \ldots \int_{x_N}^{+\infty} dy_N\, p(y_1, y_2, \ldots, y_N) \qquad (S5)$$

Then we derive probability density function. The expression of $p(x_1, x_2, \ldots, x_N)$ can be obtained by taking the mixed partial derivative of each variable of $\bar{F}(x_1, x_2, \ldots, x_N)$, we get

$$p(x_1, x_2, \ldots, x_N) = (-1)^N \frac{\partial^N}{\partial x_1 \partial x_2 \ldots \partial x_N}[\bar{F}(x_1, x_2, \ldots, x_N)] \qquad (S6)$$

For simplicity, for all variable pairs $(x_i, x_j)$ satisfying $1 \leq i < j \leq N$, define the function

$$h_{ij}(x_i, x_j) = 1, \text{ if } x_i \geq x_j \tag{S7a}$$

$$h_{ij}(x_i, x_j) = 0, \text{ if } x_i < x_j \tag{S7b}$$

Substituting

$$\max(x_i, x_j) = x_i h_{ij}(x_i, x_j) + x_j\left(1 - h_{ij}(x_i, x_j)\right) \tag{S8}$$

into Eq.(S4), we get

$$\bar{F}(x_1, x_2, \ldots, x_N) = \exp[-\sum_i r_i x_i - \sum_{i<j} r_{ij} h_{ij}(x_i, x_j) x_i - \sum_{i<j} r_{ij}(1 - h_{ij}(x_i, x_j)) x_j]$$

$$= \prod_{i=1}^{N} \exp\{-[r_i + \sum_{i<j} r_{ij} h_{ij}(x_i, x_j) + \sum_{j>i} r_{ji}(1 - h_{ji}(x_j, x_i))] x_i\} \tag{S9}$$

We only consider the probability density function when the variables $(x_1, x_2, \ldots, x_N)$ are not equal to each other ( The value of the probability density function at other points does not matter because the measure of these points is zero). Then $h_{ij}(x_i, x_j)$ are constants in the infinitesimal neighborhood of $(x_1, x_2, \ldots, x_N)$. Substituting Eq.(S9) into Eq.(S6), we have

$$p(x_1, x_2, \ldots, x_N) = (-1)^N \frac{\partial^N}{\partial x_1 \partial x_2 \ldots \partial x_N}[\bar{F}(x_1, x_2, \ldots, x_N)]$$

$$= (-1)^N \frac{\partial^N}{\partial x_1 \partial x_2 \ldots \partial x_N}\left[\prod_{i=1}^{N} \exp\{-[r_i + \sum_{i<j} r_{ij} h_{ij}(x_i, x_j) + \sum_{j>i} r_{ji}(1 - h_{ji}(x_j, x_i))] x_i\}\right]$$

$$= \prod_{i=1}^{N} (-1) \frac{\partial}{\partial x_i} \exp\{-[r_i + \sum_{i<j} r_{ij} h_{ij}(x_i, x_j) + \sum_{j>i} r_{ji}(1 - h_{ji}(x_j, x_i))] x_i\}$$

$$= \prod_{i=1}^{N} \frac{\exp\{-[r_i + \sum_{i<j} r_{ij} h_{ij}(x_i, x_j) + \sum_{j>i} r_{ji}(1 - h_{ji}(x_j, x_i))] x_i\}}{[r_i + \sum_{i<j} r_{ij} h_{ij}(x_i, x_j) + \sum_{j>i} r_{ji}(1 - h_{ji}(x_j, x_i))]}$$

$$\propto \prod_{i=1}^{N} \exp\{-[r_i + \sum_{i<j} r_{ij} h_{ij}(x_i, x_j) + \sum_{j>i} r_{ji}(1 - h_{ji}(x_j, x_i))] x_i\}$$

$$= \exp[-\sum_i r_i x_i - \sum_{i<j} r_{ij} h_{ij}(x_i, x_j) x_i - \sum_{i<j} r_{ij}(1 - h_{ij}(x_i, x_j)) x_j]$$

$$= \exp[-\sum_i r_i x_i - \sum_{i<j} r_{ij} \max(x_i, x_j)] \tag{S10}$$

Thus, the joint probability density function $p(x_1, x_2, \ldots, x_N)$ of each component lifetime satisfies

$$p(x_1, x_2, \ldots, x_N) \propto \exp[-\sum_i r_i x_i - \sum_{i<j} r_{ij} \max(x_i, x_j)] \tag{S11}$$

It can be seen that this model has a similar form to the Ising model. The first term represents the independent failure process of the component and the second term represents the coupled relationship between the two components.

Below, we explain how this model Eq.(S11) describes failure coupling between components. We analyze the failure process of a single component. For the sake of generality, we select component 1 as an example for analysis. When the lifetimes of other components are given as $X_2 = x_2, X_3 = x_3, \ldots, X_N = x_n$, the conditional probability density function of component 1 lifetime $X_1$ is

$$p(x_1|X_2 = x_2, X_3 = x_3, \ldots, X_N = x_n) = \frac{p(X_1 = x_1, X_2 = x_2, X_3 = x_3, \ldots, X_N = x_n)}{p(X_2 = x_2, X_3 = x_3, \ldots, X_N = x_n)}$$

$$= \frac{p(x_1, x_2, \ldots, x_N)}{\int_0^{+\infty} p(x_1, x_2, \ldots, x_N) dx_1} \propto \exp[-r_1 x_1 - \sum_{1<j} r_{1j} \max(x_1, x_j)] \quad (S12)$$

Considering $\max(x_1, x_j) = x_1 h_{1j}(x_1, x_j) + x_j (1 - h_{1j}(x_1, x_j))$, when $x_1$ is not equal to others, we have

$$p(x_1|X_2 = x_2, X_3 = x_3, \ldots, X_N = x_n)$$

$$\propto \exp[-r_1 x_1 - \sum_{1<j} r_{1j}[x_1 h_{1j}(x_1, x_j) + x_j (1 - h_{1j}(x_1, x_j))]]$$

$$= \exp[-r_1 x_1 - \sum_{1<j} r_{1j} x_1 h_{1j}(x_1, x_j) - \sum_{1<j} x_j (1 - h_{1j}(x_1, x_j))]$$

$$\propto \exp[-(r_1 + \sum_{1<j} r_{1j} h_{ij}(x_1, x_j)) x_1] \quad (S13)$$

Notice that there are only variable $x_1$, and $x_2, x_3, \ldots, x_N$ are all constants. It can be seen that component 1 follows an exponential distribution in this neighbor and its failure rate is

$$r_1 + \sum_{1<j} r_{1j} h_{ij}(x_1, x_j) = r_1 + \sum_{j \in U_1} r_{1j} \quad (S14)$$

Where $U_1 = \{j > 1 | x_j < x_1\}$ is set of components that failed earlier than component 1. When no component is failed yet, the failure rate of component 1 is $r_1$. If component $j$ fails earlier than component 1, then the failure rate of component 1 will increase by $r_{1j}$. Since the selection of component 1 is random, the argument here also holds for other components. The model thus describes the phenomenon where the failure of one component will lead to an increase in the failure rate of components coupled to it.

For convenience, we only consider one simple form of Eq.(S11) as

$$p(x_1, x_2, \ldots, x_N) \propto \exp\{-\beta[\sum_i x_i - \phi \sum_{i<j} A_{ij} \max(x_i, x_j)]\}, \quad (S15)$$

where β describes the time scale. $A_{ij}$ is a matrix representing the failure coupled network between components. If there is coupled failure between component $i$ and component $j$, then $A_{ij} = 1$. If there is no coupled failure between component $i$ and component $j$, then $A_{ij} = 0$. Here, we denote $A_{ij} = A_{ji}$. $\phi$ represents the component failure coupled strength. If $\phi = 0$, each component is independent and follows an exponential distribution where lifetime expectancy of $1/\beta$. If $\phi \neq 0$, a component failure will cause the failure rate of the component coupled to it increasing by $\phi\beta$. Eq. (S15) describes a failure-coupled system with network structure $A_{ij}$ and failure coupling strength $\phi$.

2. **Sample method of system component lifetime**

Since the joint probability density distribution Eq. (S15) of the component lifetime in the failure coupled network model is difficult to sample directly, the lifetimes of the network components in this study were sampled based on the Gillespie algorithm. The Gillespie algorithm is often used to efficiently and accurately simulate various stochastic processes such as chemical reactions. It is usually divided into two steps: (1) sampling the waiting time of current state; (2) sampling the next state with the transition probability to each state as the weight.

For this model, the state of the system can be defined by the currently set of functioning components $L=\{l_1, l_2, ..., l_{q(t)}\}$. During operation, the number of functioning system components decreasing continuously. When the number of failed components increased, the system switch to a new state. Therefore, the waiting time of current state the time interval from last component failure to next component failure. Then, based on the failure rate of each component, the component which is failed next time is sampled and the its lifetime value is obtained. The simulation process is as follows:

(1) Initialization. Denoted the set of functioning components as $L=\{1,2,3...,N\}$. Denoted system time as $t(0) = 0$. Here, we denote the time of $q^{th}$ components failed as $t(q)$.

(2) Calculate the failure rate of each component. If the system has $q$ failed components, the set of $N$-$q$ functioning components is $L=\{l_1, l_2, ..., l_{N-q}\}$. First calculate failure rate of each component, $\{\lambda_{l_1}, \lambda_{l_2}, ..., \lambda_{l_{N-q}}\}$. By equation (S14-S15), the failure rate of functional components $i \in \{l_1, l_2, ..., l_{N-q}\}$ is

$$\lambda_i = \beta[1 + \sum_j A_{ij} d_j(t) \times \phi], \tag{S16}$$

where $d_j(t)$=0 means that component $j$ is alive at time $t$. $d_j(t)$=1 indicates that component $j$ failed at time $t$. $\sum_j A_{ij} d_j(t(q))$ represents the number of failed neighbors for component $i$ at time $t(q)$.

(3) Calculate the total failure rate. The probability ($\lambda_{tot}$) that a network component fails per

unit time currently can be obtained based on the failure rate of each functioning component $\lambda_i (i \in \{l_1, l_2, ..., l_{N-q}\})$. Considering infinitely short time $dt$, the probability that system component $i$ failed is $\lambda_i dt$. Thus, the probability that at least one component fails in this infinitely short time $(t, t+dt)$ is

$$\lambda_{tot} dt = 1 - \prod_{i \in \{l_1, l_2, ..., l_{N-q}\}} (1 - \lambda_i dt) = \sum_{i \in \{l_1, l_2, ..., l_{N-q}\}} \lambda_i \, dt + o(dt) \quad (S17)$$

Here, the probability that two or more components fail at the same time in $dt$ time is a high-order quantity about $dt$, which can be ignored. Therefore, two components will not fail at the same time. Thus, the probability that a component fails in the network per unit time is:

$$\lambda_{tot} = \sum_{i \in \{l_1, l_2, ..., l_{N-q}\}} \lambda_i \quad (S18)$$

(4) Sample time interval $\tau_q$ from the moment $t(q)$ to $t(q+1)$ when $(q+1)^{th}$ component failure happen. This time interval $\tau_q$ follows exponential distribution as

$$p(\tau) = \lambda_{tot} \exp(-\lambda_{tot} \tau) \quad (S19)$$

We sample $\tau_q$ from this exponential distribution. Then we can obtain the moment when the $(q+1)^{th}$ component in system as

$$t(q+1) = t(q) + \tau_q \quad (S20)$$

(5) Sample the component $l_{fail}$ ($l_{fail} \in \{l_1, l_2, ..., l_{N-q}\}$) which failed at the moment $t(q+1)$ and record the lifetime of this component as $LifeTime[l_{fail}]$. According to equation (S17), the probability that each component fails at $t(q+1)$ is proportional to its own failure rate. Therefore, choosing $\lambda_{l_1}, \lambda_{l_2}, ... \lambda_{l_q}$ as the weight of each component failure, sample the component $l_{fail}$ fail at $t(q+1)$ from the set of functioning components $\{l_1, l_2, ..., l_{q(t)}\}$. Then we denoted the lifetime of $l_{fail}$ as

$$LifeTime[l_{fail}] = t(q+1) \quad (S21)$$

(6) Update the current time to $t = t(q+1)$ and remove component $l_{fail}$ from the set of functioning components $L$.

(7) If there are still functioning components in set $L$, repeat steps (1) through (5). If not, the sample is completed and the lifetime values of all components in the network are obtained.

3. **Mortality calculation method**

When calculating the system mortality, we divide the interval [0,2] into 400 bins when failure coupling strength is 0, and into 100 bins when coupling strength is $10^4$ or $10^6$. For $i^{th}$ interval $[t_i, t_{i+1}]$, we calculate the number of alive samples at $t_i$ time and $t_{i+1}$ time, denoted as $n_i$ and $n_{i+1}$. The mortality value $t = \frac{t_i+t_j}{2}$ is calculated by

$$\lambda\left(\frac{t_i+t_j}{2}\right) = \frac{n_{i+1}-n_i}{n_{i+1}(t_{i+1}-t_i)}. \tag{S22}$$

## 4. KL divergences calculation method between fitting model and empirical data

We discretize the data to calculate the KL divergence between the model distribution and the empirical distribution. For given lifetime samples, we divide the lifetime interval ([minimum life time, maximum liftime]) into 50 bins, denoted as $[t'_0, t'_1), [t'_1, t'_2), ..., [t'_{49}, t'_{50}]$. We calculate the experience probability $Q_i$ that samples are in $i^{th}$ interval $[t'_{i-1}, t'_i]$. The probability density function corresponding to the distribution model obtained by the maximum likelihood method is denoted as $f(a)$. The KL divergence is obtained by the following equation:

$$KL(Q|f) = -\sum_i Q_i \ln\left(\frac{Q_i}{\int_{t'_{i-1}}^{t'_i} f(a)da}\right)$$

If no sample is in $j^{th}$ interval making experience probability $Q_j = 0$, the limit $\lim_{Q_j \to 0} Q_j \ln Q_j = 0$ is used.

## 5. The relationship between the lifetime distribution and mortality of the system over time with a fixed system size in ER networks, small world network and BA network.

The relationship between the lifetime distribution and mortality of the system over time with a fixed system size in ER networks, small world network and BA network are shown in Fig. S2, Fig. S3 and Fig. S4. System size is 6400. The fitting effect is measured by calculating the KL divergence between the fitting function and the simulation data. KL divergences of three fitting model and empirical data under different coupled strength are shown in Table S1, S2 and S3.

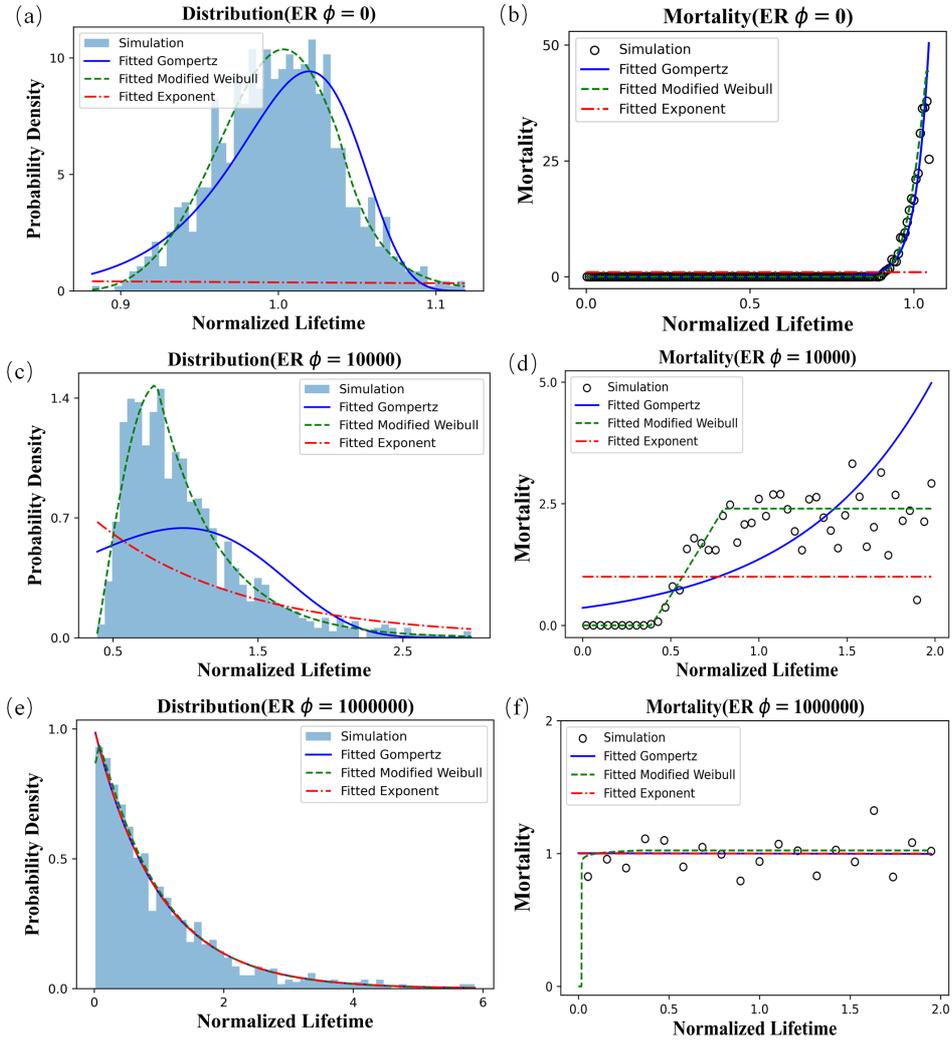

**Fig. S2 The relationship between the lifetime distribution and mortality of the system over time with different failure coupling strength in ER network.** (a), (c) and (e) are the lifetime distribution of the system when the coupled strength is 0, $10^4$, and $10^6$ respectively. (b), (d) and (f) are the system mortality over time when the coupled strength is 0, $10^4$, and $10^6$ respectively. The blue solid, green dashed and red solid dotted fitting curves in the figure are the Gompertz model, modified Weibull distribution and exponential distribution respectively. The average degree of ER network is 4.

Table S1 KL divergences of three fitting model and empirical data under different coupled strength in ER network

|  | Gompertz model | modified Weibull distribution | exponential distribution |
|---|---|---|---|
| $\phi=0$ | 0.102 | 0.0224 | 2.8505 |
| $\phi=10^4$ | 0.3728 | 0.03 | 0.6865 |
| $\phi=10^6$ | 0.0396 | 0.0214 | 0.0396 |

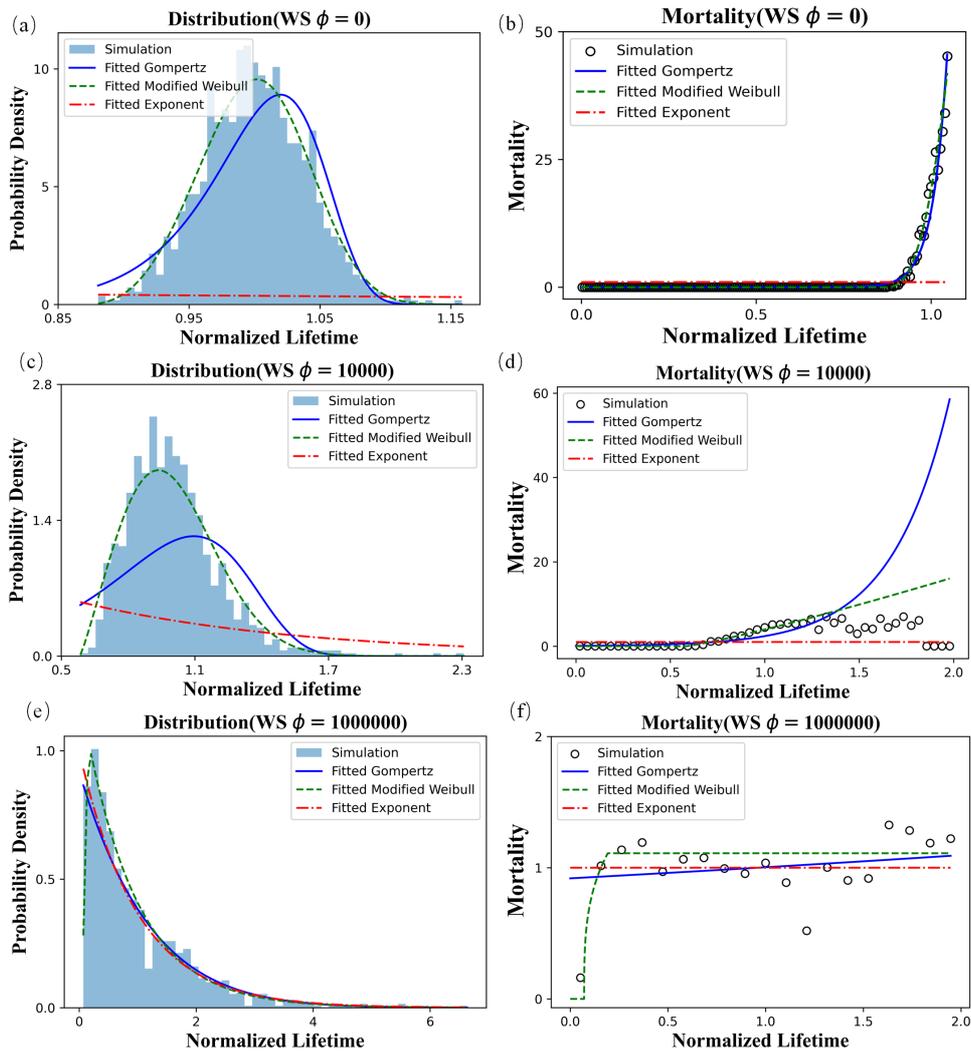

**Fig. S3 The relationship between the lifetime distribution and mortality of the system over time with different failure coupling strength in small world network.** (a), (c) and (e) are the lifetime distribution of the system when the coupled strength is 0, $10^4$, and $10^6$ respectively. (b), (d) and (f) are the system mortality over time when the coupled strength is 0, $10^4$, and $10^6$ respectively.

The blue solid, green dashed and red solid dotted fitting curves in the figure are the Gompertz model, modified Weibull distribution and exponential distribution respectively. The average degree of the network is 4 and the rewiring probability is 0.01 during constructing the small world network.

Table S2 KL divergences of three fitting model and empirical data under different coupled strength in small world network

|  | Gompertz model | modified Weibull distribution | exponential distribution |
|---|---|---|---|
| $\emptyset=0$ | 0.1025 | 0.0143 | 2.8255 |
| $\emptyset=10^4$ | 0.3766 | 0.0429 | 1.2913 |
| $\emptyset=10^6$ | 0.1007 | 0.0277 | 0.1055 |

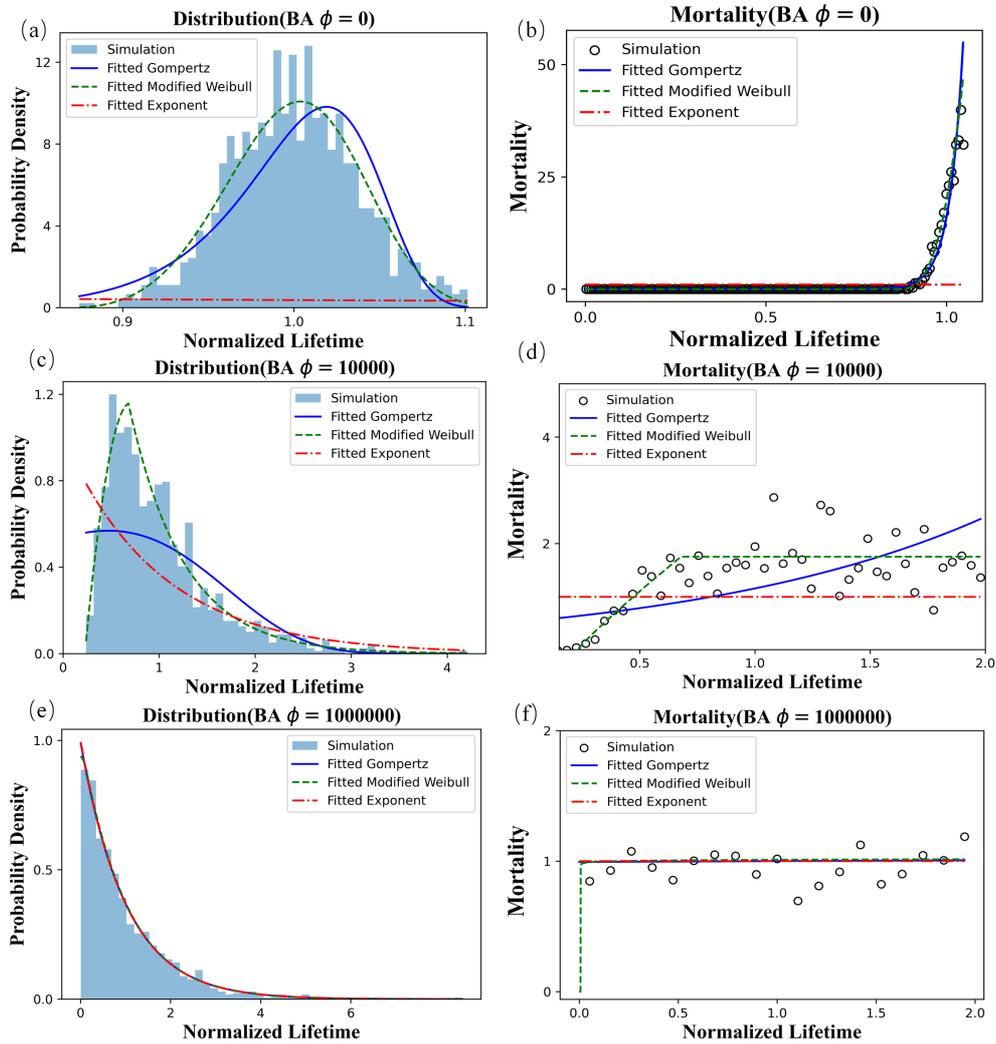

**Fig. S4 The relationship between the lifetime distribution and mortality of the system over time with different failure coupling strength in BA network.** (a), (c) and (e) are the lifetime distribution of the system when the coupled strength is 0, $10^4$, and $10^6$ respectively. (b), (d) and (f) are the system mortality over time when the coupled strength is 0, $10^4$, and $10^6$ respectively. The blue solid, green dashed and red solid dotted fitting curves in the figure are the Gompertz model, modified Weibull distribution and exponential distribution respectively. The average degree of BA network is 4.

Table S3 KL divergences of three fitting model and empirical data under different coupled strength in BA network

|  | Gompertz model | modified Weibull distribution | exponential distribution |
|---|---|---|---|
| $\emptyset=0$ | 0.1024 | 0.0275 | 2.8833 |
| $\emptyset=10^4$ | 0.2585 | 0.029 | 0.415 |
| $\emptyset=10^6$ | 0.018 | 0.0113 | 0.0181 |

6. **The relationship between the lifetime distribution and mortality of the system over time with fixed failure coupled strength in ER networks, small world network and BA network.**

The relationship between the lifetime distribution and mortality of the system over time with fixed failure coupled strength in ER networks, small world network and BA network are shown in Fig. S5, Fig. S6 and Fig. S7. Coupled strength is $10^4$. The fitting effect is measured by calculating the KL divergence between the fitting function and the simulation data. KL divergences of three fitting model and empirical data under different coupled strength are shown in Table S4, S5 and S6.

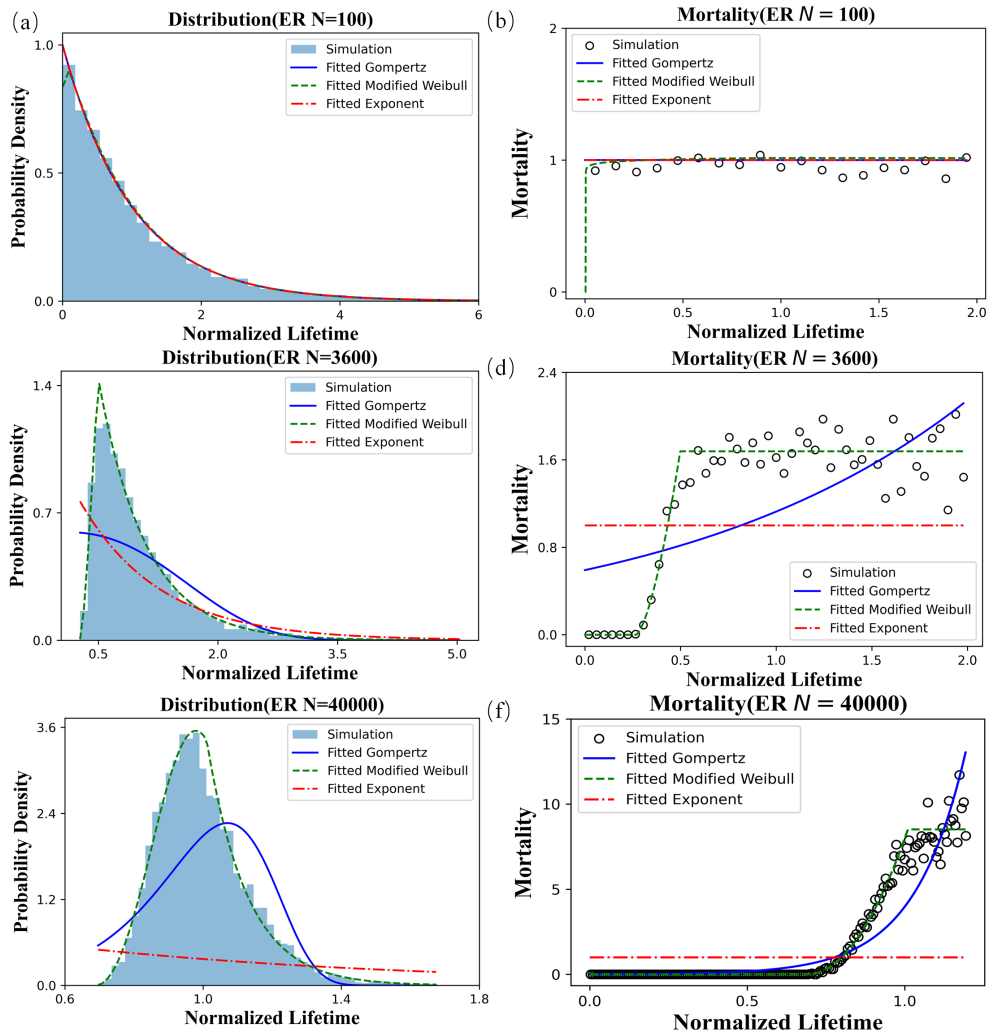

**Fig. S5 The relationship between the lifetime distribution and mortality of the system over time with different system sizes in the ER network**. (a), (c) and (e) are the lifetime distribution of the system when the system size is 100, 3600 and 40000 respectively. (b), (d) and (f) are system mortality over time when the system size is 100, 3600, and 40000, respectively. The blue solid, green dashed and red solid dotted fitting curves in the figure are the Gompertz model, modified Weibull distribution and exponential distribution respectively. The average degree of ER network is 4.

Table S4 KL divergences of three fitting model and empirical data in different system size in ER network

|          | Gompertz model | modified Weibull distribution | exponential distribution |
|----------|----------------|-------------------------------|--------------------------|
| N=100    | 0.0144         | 0.0103                        | 0.0146                   |
| N=3600   | 0.3061         | 0.0141                        | 0.4416                   |
| N=40000  | 0.18           | 0.0324                        | 1.6589                   |

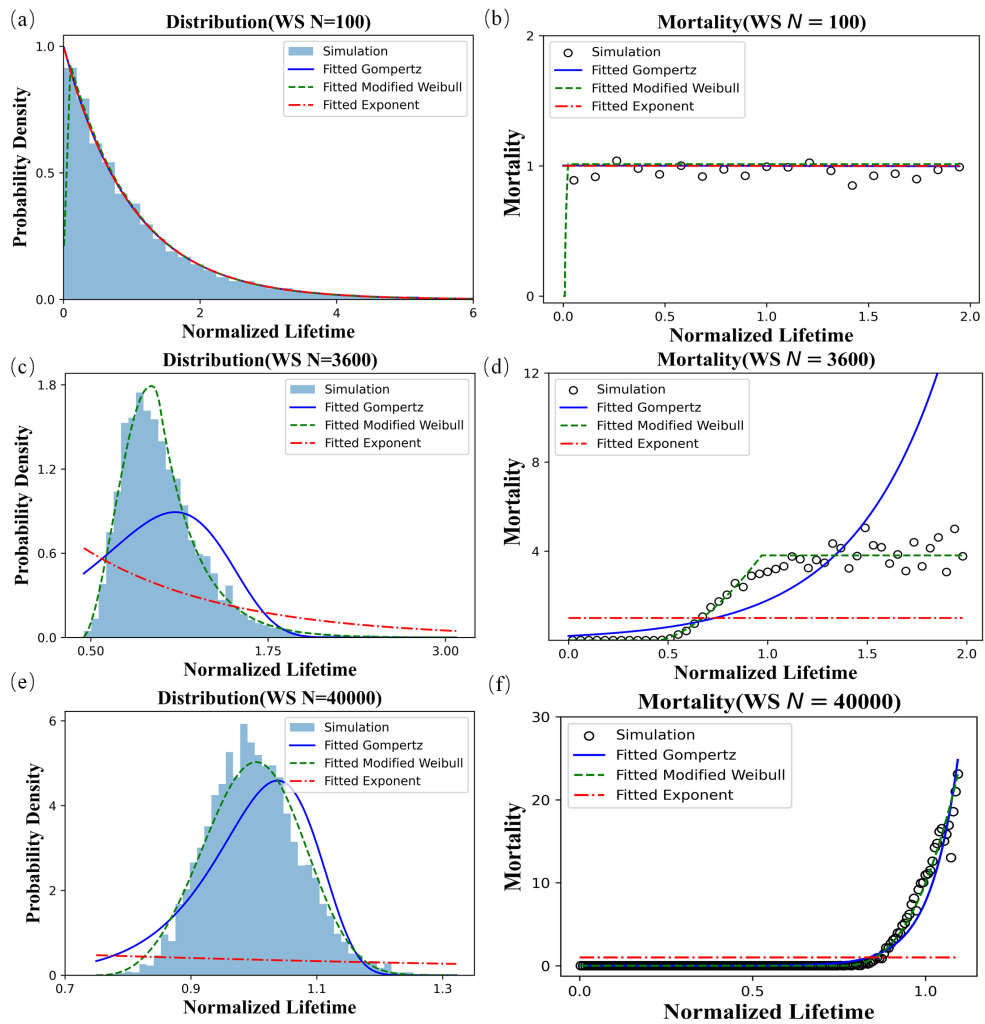

**Fig. S6 The relationship between the lifetime distribution and mortality of the system over time with different system sizes in the small world network**. (a), (c) and (e) are the lifetime distribution of the system when the system size is 100, 3600 and 40000 respectively. (b), (d) and (f) are system mortality over time when the system size is 100, 3600, and 40000, respectively. The blue solid, green dashed and red solid dotted fitting curves in the figure are the Gompertz model,

modified Weibull distribution and exponential distribution respectively. The average degree of the network is 4 and the rewiring probability is 0.01 during constructing the small world network.

Table S5 KL divergences of three fitting model and empirical data in different system size in small world network

|  | Gompertz model | modified Weibull distribution | exponential distribution |
|---|---|---|---|
| N=100 | 0.0168 | 0.0079 | 0.0167 |
| N=3600 | 0.3572 | 0.0121 | 0.9599 |
| N=40000 | 0.139 | 0.0176 | 2.2126 |

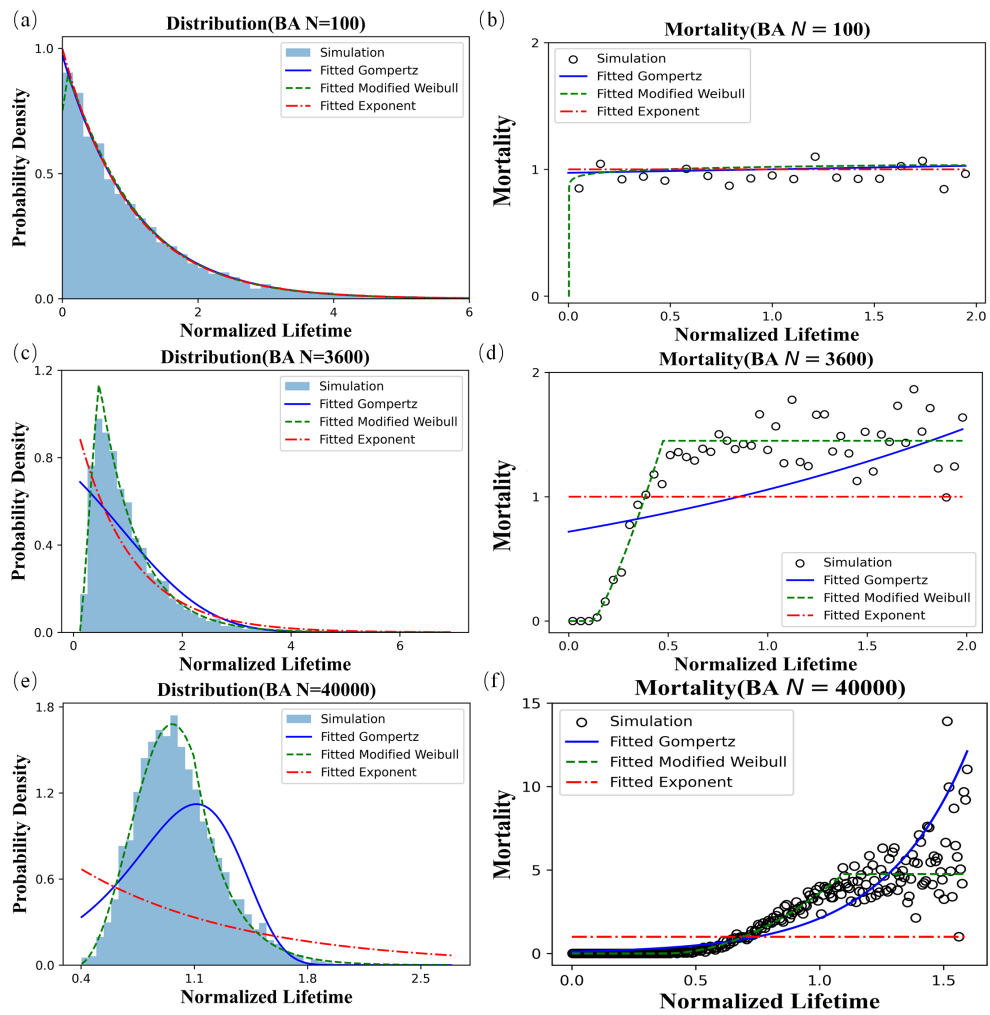

**Fig. S7 The relationship between the lifetime distribution and mortality of the system over time with different system sizes in the BA network**. (a), (c) and (e) are the lifetime distribution of the system when the system size is 100, 3600 and 40000 respectively. (b), (d) and (f) are system mortality over time when the system size is 100, 3600, and 40000, respectively. The blue solid, green dashed and red solid dotted fitting curves in the figure are the Gompertz model, modified Weibull distribution and exponential distribution respectively. The average degree of BA network is 4.

Table S5 KL divergences of three fitting model and empirical data in different system size in BA network

|  | Gompertz model | modified Weibull distribution | exponential distribution |
| --- | --- | --- | --- |
| N=100 | 0.0147 | 0.0118 | 0.016 |
| N=3600 | 0.196 | 0.0117 | 0.265 |
| N=40000 | 0.1979 | 0.0206 | 1.0076 |

7. **Phase diagram of network lifetime distribution in ER networks, small world network and BA network**.

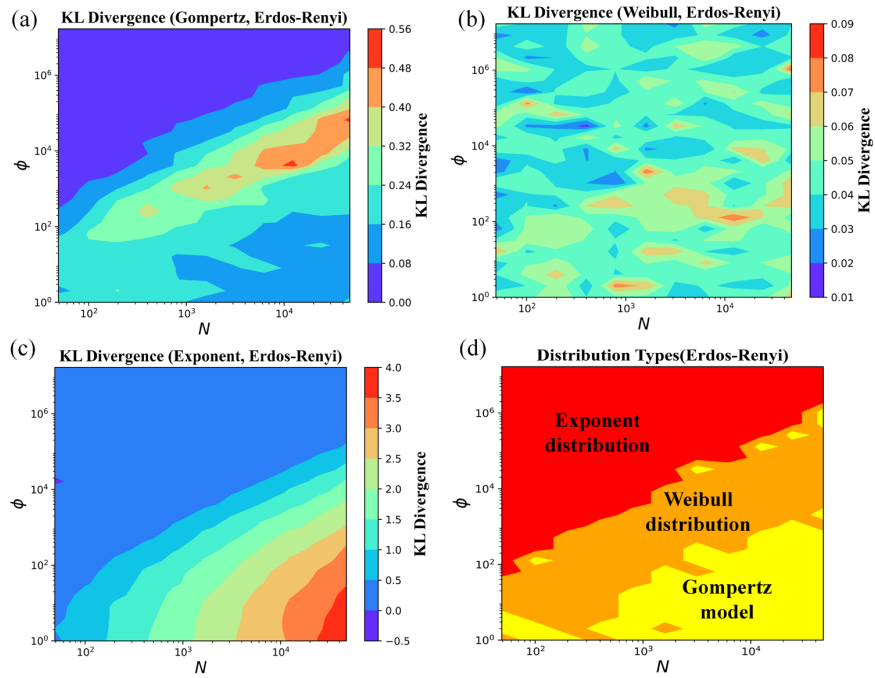

**Fig. S8 Phase diagram of network lifetime distribution in ER networks.** (a)-(c) KL Divergence between empirical data and Gompertz model, modified Weibull distribution model and Exponent distribution model under different coupled strengths and different system sizes. (d) ER network lifetime distribution types under different coupled strengths and different system sizes. The average degree of ER network is 4.

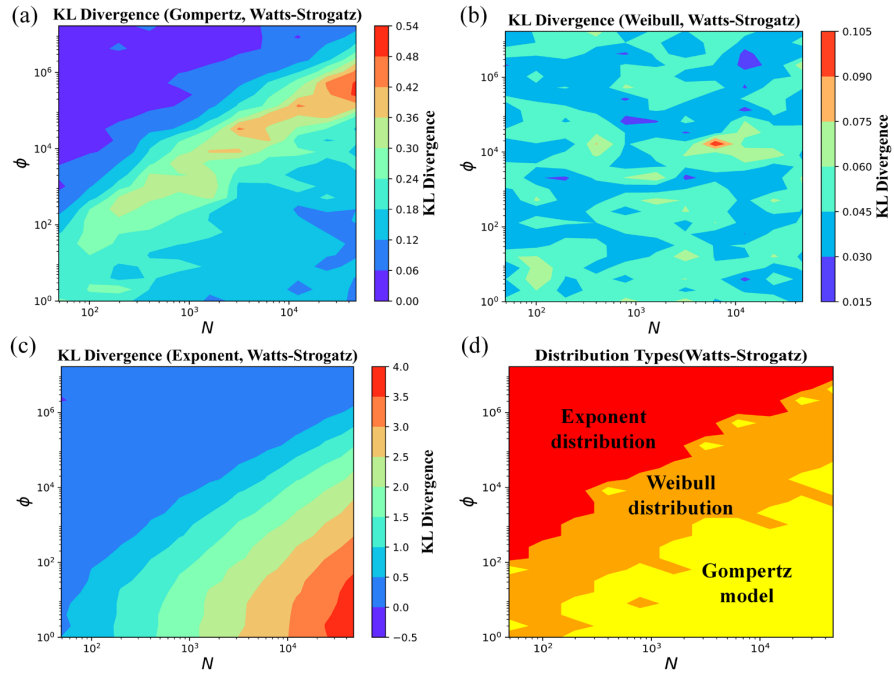

**Fig. S9 Phase diagram of network lifetime distribution in small world networks.** (a)-(c) KL Divergence between empirical data and Gompertz model, modified Weibull distribution model and Exponent distribution model under different coupled strengths and different system sizes. (d)Small world network lifetime distribution types under different coupled strengths and different system sizes. The average degree of the network is 4 and the rewiring probability is 0.01 during constructing the small world network.

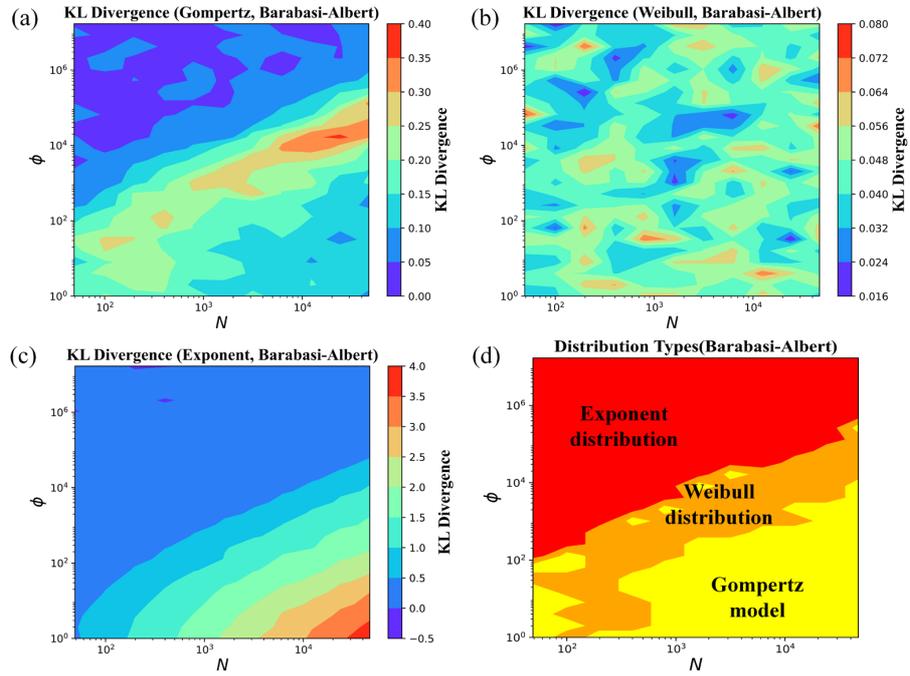

**Fig. S10 Phase diagram of network lifetime distribution in BA networks.** (a)-(c) KL Divergence between empirical data and Gompertz model, modified Weibull distribution model and Exponent distribution model under different coupled strengths and different system sizes. (d)BA network lifetime distribution types under different coupled strengths and different system sizes. The average degree of BA network is 4.

8. **The KL divergence values between the Gompertz model and empirical distribution under different system sizes and failure coupling strength in ER network, small-world network and BA network.**

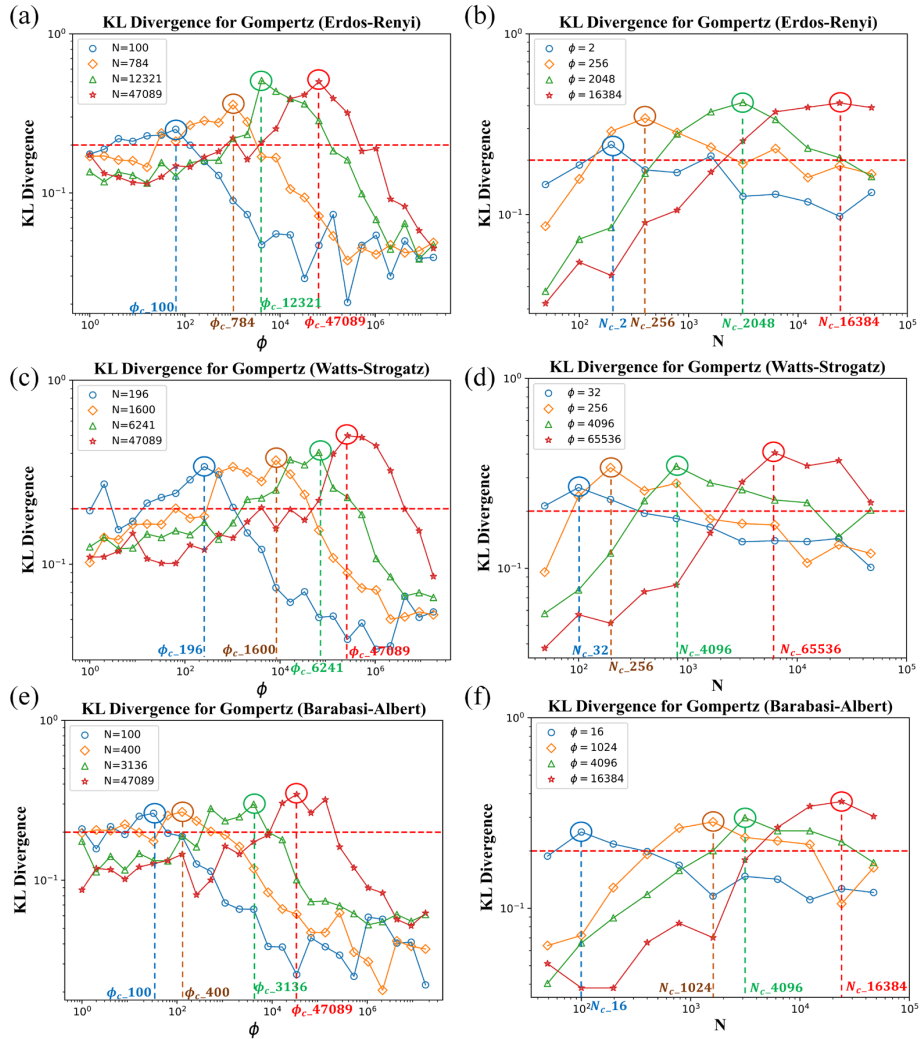

**Fig. S11 The KL divergence values between the Gompertz model and empirical distribution under different system sizes and failure coupling strength in ER network, small-world network and BA network.** (a), (c) and (e) are the variation of KL divergence values between the Gompertz model and empirical distribution with failure coupling strength when the system size is fixed in ER network, small-world network and BA network, respectively. (b), (d) and (e ) are the variation of KL divergence values between the Gompertz model and empirical distribution with system size when the failure coupling strength is fixed in ER network, small-world network and BA network, respectively.